\documentclass[aps,prb,reprint,twocolumn,showpacs,floatfix,superscriptaddress,nofootinbib]{revtex4}
\usepackage{graphicx}
\usepackage{amssymb}
\usepackage{amsmath}
\usepackage{lmodern}
\usepackage{color}
\usepackage{hyperref}
\usepackage[T2A,T1]{fontenc}
\usepackage[utf8]{inputenc}
\usepackage[russian,french,english]{babel}
\usepackage{epstopdf}
\begin{document}

\title{RKKY interaction in graphene} 

\author{E. Kogan} \email{kogan@biu.ac.il}

\affiliation{Department of Physics, Bar-Ilan University, Ramat-Gan 52900, Israel} \date{\today}

\begin{abstract} We consider RKKY interaction between two magnetic impurities in graphene. The consideration is based on the perturbation theory for the thermodynamic potential in the imaginary time representation. We analyze the symmetry of the RKKY interaction on the bipartite lattice at half filling. Our  analytical calculation of the interaction is based on direct evaluation of real space spin susceptibility.
We show in the Appendix, added to the published version, that the approach can be easily generalized to the case of finite temperature.
\end{abstract}


\maketitle

\section{Introduction}

Since   graphene   was first isolated experimentally \cite{novoselov},  it is in the focus of attention of both   theorists and experimentalists.  Many physical phenomena, well studied in "traditional" solid state physics  look quite different in graphene. In this paper we will talk about the Ruderman--Kittel--Kasuya--Yosida (RKKY) interaction,  first studied (in a normal metal) more than 60 years ago \cite{ruderman,kasuya,yosida}.
 This interaction
 is the effective  exchange  between two magnetic impurities in a non--magnetic host, obtained  as the second order  perturbation  with respect to exchange interaction between the
magnetic impurity and the itinerant electrons of the host.

Quite a few theoretical papers published recently considered RKKY interaction in graphene
 \cite{vozmediano,dugaev,brey,saremi,black,sherafati,uchoa}.
 Though analysis of the RKKY interaction is simple in principle,
 calculation of the integrals defining the  interaction (whether analytical or numerical)
can pose some problems. However, substantial progress was achieved in the field.

Our interest in  the RKKY interaction in graphene started from learning about the theorem stating that for any half--filled bipartite lattice the exchange interaction between the magnetic adatoms is ferromagnetic, if the adatoms belong to the same sublattice, and antiferromagnetic, if the adatoms belong to  different sublattices \cite{saremi}. Also, in this paper and in the following one \cite{sherafati}, in the approximation of the  linear dispersion law for the electrons, the RKKY interaction in graphene was calculated analytically.
 However, the integrals obtained in both papers turned out to be divergent, and the complicated (and to some extent arbitrary) cut-off procedure was implemented to obtain from these integrals the finite results. So we started to look for the procedure which will allow
to eliminate this problem. The second reason for our interest was the fact that the theorem, mentioned above, was challenged \cite{bunder}. The claim was that the proof is based on  calculation of the magnetic susceptibility
 of the free electron gas by the imaginary--time method,  demanding later analytic continuation from the imaginary frequencies to the real ones.
 On the other hand, consideration by the real--time method, presented in Ref. \cite{bunder} does not support the statement of the theorem.
To clarify the situation and get rid of the shortages mentioned above, we decided to analyze the problem of RKKY interaction in graphene from the scratch.

\section{RKKY interaction}

We consider two magnetic impurities at the sites $i$ and $j$ and  assume a contact exchange interaction between the electrons and the magnetic impurities. Thus the total Hamiltonian of the system is \begin{eqnarray} \label{intera0} H_T=H+H_{int}=H-J{\bf S}_i{\bf\cdot s}_i-J{\bf S}_j{\bf\cdot s}_j, \end{eqnarray} where $H$ is the Hamiltonian of the electron system, ${\bf S}_i$  is the spins of the impurity and ${\bf s}_i$ is the spin of itinerant electrons at site $i$.

Our consideration  is based on the perturbation theory for the thermodynamic potential \cite{abrikosov}. The correction to the thermodynamic potential due to interaction is \begin{eqnarray} \Delta\Omega=-T\ln\left\langle S \right\rangle\equiv -T\ln {\rm tr}\left\{S\cdot e^{-H/T}/Z\right\}, \end{eqnarray} where the $S$--matrix is given by the equation \begin{eqnarray} S=\exp\left\{-\int_0^{1/T} H_{int}(\tau)d\tau\right\}. \end{eqnarray} Writing down ${\bf s}_i$ in the second quantization representation \begin{eqnarray} {\bf s}_i=\frac{1}{2}c_{i\alpha}^{\dagger}{\bf\sigma}_{\alpha\beta}c_{i\beta}, \end{eqnarray} the second order term of the expansion with respect to the interaction is \begin{eqnarray} \label{omega} &&\Delta\Omega=\frac{J^2T}{4}\sum_{\alpha\beta\gamma\delta}{\bf S}_i{\bf\cdot \sigma}_{\alpha\beta}{\bf S}_j{\bf\cdot \sigma}_{\gamma\delta}\\ &&\int_0^{1/T}\int_0^{1/T}d\tau_1d\tau_2 \left\langle T_{\tau}\left\{ c_{i\alpha}^{\dagger}(\tau_1)c_{i\beta}(\tau_1) c_{j\gamma}^{\dagger}(\tau_2)c_{j\delta}(\tau_2)\right\} \right\rangle.\nonumber \end{eqnarray} Notice that we have ignored the terms proportional to ${\bf S}_i^2$ and ${\bf S}_j^2$, because they are irrelevant for our calculation of the effective interaction between the adatoms spins.

Leaving aside the question about the spin structure of the two--particle Green's function standing in the r.h.s. of Eq. (\ref{omega}) (for interacting electrons), further on we assume that the electrons are non--interacting. This will allow us to use Wick theorem and present the correlator from Eq. (\ref{omega}) in the form \begin{eqnarray} -{\cal G}_{\beta\gamma}(i,j;\tau_1-\tau_2){\cal G}_{\delta\alpha}(j,i;\tau_2-\tau_1), \end{eqnarray} where \begin{eqnarray} {\cal G}_{\beta\gamma}(i,j,\tau_1-\tau_2)=-\left\langle T_{\tau}\left\{ c_{i\beta}(\tau_1) c_{j\gamma}^{\dagger}(\tau_2)\right\} \right\rangle \end{eqnarray} is the Matsubara Green's function \cite{abrikosov}. We can connect ${\cal G}_{\beta\gamma}$ with the Green's function of spinless electron \begin{eqnarray} {\cal G}_{\beta\gamma}(i,j,\tau_1-\tau_2)=-\delta_{\beta\gamma}\left\langle T_{\tau}\left\{ c_{i}(\tau_1) c_{j}^{\dagger}(\tau_2)\right\} \right\rangle. \end{eqnarray} Presence of delta-symbols allows to perform summation with respect to spin indices in Eq. (\ref{omega}) \begin{eqnarray} \sum_{\alpha\beta}{\bf S}_i{\bf\cdot \sigma}_{\alpha\beta}{\bf S}_j{\bf\cdot \sigma}_{\beta\alpha}={\bf S}_i{\bf\cdot  S}_j, \end{eqnarray} which gives \begin{eqnarray} \Delta\Omega=-J^2\chi_{ij}{\bf S}_i{\bf\cdot  S}_j, \end{eqnarray} where \begin{eqnarray} \label{abr} \chi_{ij}=-\frac{1}{4}\int_{0}^{1/T}{\cal G}(i,j;\tau){\cal G}(j,i;-\tau)d\tau \end{eqnarray} is the free electrons static real space spin susceptibility.

Thus we obtain \begin{eqnarray} H_{RKKY}=-J^2\chi_{ij}{\bf S}_i{\bf\cdot  S}_j, \end{eqnarray} Eq. (\ref{abr}) was applied to calculation of RKKY interaction in graphene for the first time, to the best of our knowledge, in Ref. \cite{cheianov}.

The Green's function can be easily written down using representation of eigenvectors and eigenvalues of the operator $H$ \begin{eqnarray} \label{eigen} \left(H-E_n\right)u_n=0. \end{eqnarray} It is \begin{eqnarray} \label{abr2} {\cal G}(i,j;\tau)=\sum_n u_n^*(i)u_n(j)e^{-\xi_n\tau}\nonumber\\ \times\left\{\begin{array}{ll}-\left(1-n_F(\xi_n)\right), &\tau>0\\ n_F(\xi_n), & \tau<0 \end{array} \right., \end{eqnarray} where $\xi_n=E_n-\mu$, and  $n_F(\xi)=\left(e^{\beta\xi}+1\right)^{-1}$ is the Fermi distribution function.

\section{Symmetry of the RKKY interaction on the half-filled bipartite lattice}

In this Section we'll consider the Hamiltonian of the free electrons in tight-binding representation \begin{eqnarray} H=\sum_{i,j}t_{ij}c_{i}^{\dagger}c_{j}. \end{eqnarray} Bipartite lattice we'll understand in the sense, that all the sites can be divided in two sublattices, and there is only inter--sublattice hopping (no intra--sublattice hopping). Thus the Hamiltonian $H$ in matrix representation is \begin{eqnarray} \label{bp} H=\left(\begin{array}{cc} 0 & T \\ T^{\dagger} & 0 \end{array}\right), \end{eqnarray} where $T$ is some  matrix $N\times M$ ( the first $N$ sites belong to the sublattice $A$ and the last $M$ sites belong to sublattice $B$).

Consider a matrix of even  more general form than (\ref{bp}) \begin{eqnarray} \label{bp3} \tilde{H}=\left(\begin{array}{cc} 0_{N\times N} & B_{N\times M} \\ C_{M\times N} & 0_{M\times M} \end{array}\right); \end{eqnarray} $B$ and $C$ are some arbitrary matrices. The spectrum of the matrix $\tilde{H}$ can be found from a secular equation \begin{eqnarray} \label{bp4} \left|\begin{array}{cc} -EI_{N\times N} & B_{N\times M}  \\ C_{M\times N} & -EI_{M\times M}  \end{array}\right|=0. \end{eqnarray} In Ref. \cite{gantmacher} it is proved the following property of the determinant of the block matrix \begin{eqnarray} \label{gan} \left|\begin{array}{cc} A_{N\times N} & B_{N\times M} \\ C_{M\times N} & D_{M\times M} \end{array}\right|=\left| A \right|\left|D- C A^{-1} B\right|, \end{eqnarray} which is valid, provided $|A|\neq 0$. For non-zero eigenvalues of the matrix $\tilde{H}$, we can apply Eq. (\ref{gan}) to the determinant (\ref{bp4}) to get \begin{eqnarray} \label{gan2} \left| E^2I_{M\times M} - C  B\right|=0. \end{eqnarray} Thus the spectrum of the bipartite Hamiltonian is symmetric, that is non-zero eigenvalues of the matrix $H$  are present in pares $(E,-E)$.

If we write down Eq. (\ref{eigen}) explicitly in a matrix form \begin{eqnarray} \left(\begin{array}{cc}-E_n I & T\\T^{\dagger} & -E_nI\end{array} \right)u_n=0, \end{eqnarray} it becomes obvious that \begin{eqnarray} \label{pm} u_{\bar{n}}(i)=\pm u_n(i), \end{eqnarray} where $u_n$ is the eigenfunction corresponding to $E_n$ and $u_{\bar{n}}$ is the eigenfunction corresponding to $-E_n$, and in the r.h.s. of Eq. (\ref{pm}) there is plus sign if the site $i$ belongs to one sublattice, and there is minus sign if the site belongs to the opposite sublattice.

Eq.(\ref{pm}) and the fact that for $\mu=0$ we have $n_F(\xi_{\bar m})=1-n_F(\xi_m)$, immediately convince us that the terms with non-zero energy in Eq. (\ref{abr2}) are pairwise antisymmetric (with respect to simultaneous transformation $\tau\to -\tau$, $i\rightleftarrows j$ and complex conjugation) for the sites $i$ and $j$ belonging to the same sublattice, and pairwise symmetric for the sites $i$ and $j$ belonging to opposite sublattices. The term (terms) with $E=0$ is antisymmetric with respect to the above mentioned transformation, no matter which sublattices the sites belong to. Thus for the sites $i$ and $j$ belonging to the same sublattice \begin{eqnarray} \label{simsim} {\cal G}(j,i;-\tau)=-{\cal G}^*(i,j;\tau). \end{eqnarray} For the sites $i$ and $j$ belonging to different sublattices \begin{eqnarray} \label{simsim2} {\cal G}(j,i;-\tau)={\cal G}^*(i,j;\tau), \end{eqnarray} provided there are no zero energy states, or we can neglect there contribution to the Green's function.

Thus
 for the case considered, Eq. (\ref{abr}) gives  ferromagnetic exchange between magnetic impurities on the same sublattice and
 antiferromagnetic exchange between impurities on  opposite sublattices (under the restriction presented above).

\section{Analytic calculation of the RKKY interaction in graphene}

In  calculations of the RKKY interaction in graphene the $\sum_n$ in Eq. (\ref{abr2}) turns into $\frac{a^2}{(2\pi)^2}\int d^2{\bf p}$, where $a$ is the carbon--carbon distance. (Actually, there should appear a numerical multiplier, connecting the area of the elementary cell with $a^2$, but we decided to discard it, which is equivalent to some numerical renormalization of $J$.) Also
\begin{eqnarray}
\label{u}
u_n(i)=e^{i{\bf p\cdot R}_i}\psi_{\bf p},
\end{eqnarray} where $\psi_{\bf p}$
is the appropriate component of spinor electron wave-function (depending upon which sublattice the magnetic adatom belongs to) in momentum representation.

Further on the integration with respect to $d^2{\bf p}$ we'll treat as the integration in the vicinity of two Dirac points $K,K'$ and present ${\bf p}={\bf K}({\bf K}')+{\bf k}$. The wave function for the momentum around Dirac points $K$ and $K'$ has respectively the form \begin{eqnarray} \psi_{\nu,{\bf K}}({\bf k})=\frac{1}{\sqrt{2}}\left(\begin{array}{l}e^{-i\theta_{\bf k}/2}\\ \nu e^{i\theta_{\bf k}/2}\end{array}\right)\nonumber\\ \psi_{\nu,{\bf K}'}({\bf k})=\frac{1}{\sqrt{2}}\left(\begin{array}{l}e^{i\theta_{\bf k}/2}\\ \nu e^{-i\theta_{\bf k}/2}\end{array}\right), \end{eqnarray} where $\nu=\pm 1$ corresponds to electron and hole band
   \cite{castro};  the upper line of the spinor refers to the sublattice $A$ and the lower line refers to the sublattice $B$.

The  chemical potential is at the Dirac points; $E_+({\bf k})$ and  $E_-({\bf k})$ would be electron and hole energy. Then   Eq. (\ref{abr2})
  takes the form:  for $i$ and $j$ belonging to the same sublattice
\begin{eqnarray}
\label{rk}
{\cal G}^{AA}(i,j;\tau>0)=-\frac{1}{2}\frac{a^2}{(2\pi)^2}\int d^2{\bf k} e^{i{\bf k \cdot R}_{ij}-E_+({\bf k})\tau}\nonumber\\
\left[e^{i{\bf K \cdot R}_{ij}}+e^{i{\bf K' \cdot R}_{ij}}\right],
\end{eqnarray}
and for $i$ and $j$ belonging to  different sublattices
\begin{eqnarray}
\label{rk2}
&&{\cal G}^{AB}(i,j;\tau>0)=\frac{1}{2}\frac{a^2}{(2\pi)^2}\int d^2{\bf k} e^{-E_+({\bf k})\tau}\nonumber\\
&&\times\left[ e^{i({\bf K}+{\bf k}){\bf \cdot R}_{ij}- i\theta_k}-e^{i({\bf K}'+{\bf k}){\bf \cdot R}_{ij}+ i\theta_k}\right].
\end{eqnarray}
For $\tau<0$ we should change the sign of the Green's functions and substitute $E_-$ for $E_+$.

For isotropic dispersion law $E({\bf k})=E(k)$ we can perform   the angle integration in Eqs. (\ref{rk})  and (\ref{rk2}) to get
\begin{eqnarray}
\label{aaa4}
&&\frac{1}{2\pi}\int d^2{\bf k}e^{i{\bf k \cdot R}_{ij}-E(k)\tau} =\int_0^{\infty}dk kJ_0(kR)e^{-E(k)\tau} \nonumber\\
&&\frac{1}{2\pi}\int d^2{\bf k}e^{i{\bf k \cdot R}_{ij}\pm i\theta_k-E(k)\tau} \\\nonumber
&&=e^{\pm i\theta_{\bf R}}\int_0^{\infty}dk kJ_1(kR)e^{-E(k)\tau}
\end{eqnarray}
($J_0$ and $J_1$ are the Bessel function of zero and first order respectively, and $\theta_{\bf R}$ is the angle between the vectors ${\bf K}-{\bf K}'$ and ${\bf R}_{ij}$).

For the linear  dispersion law
\begin{eqnarray}
E_{\pm}(k)=\pm v_Fk,
\end{eqnarray} using mathematical identity \cite{prudnikov}
\begin{eqnarray}
\label{identity}
&&\int _0^{\infty}x^{n-1}e^{-px}J_{\nu}(cx)dx\\ &&=(-1)^{n-1}c^{-\nu}\frac{\partial^{n-1}}{\partial p^{n-1}}\frac{\left(\sqrt{p^2+c^2}-p\right)^{\nu}} {\sqrt{p^2+c^2}},\nonumber
\end{eqnarray}
we can  explicitly perform the remaining integration.  Calculating  integrals (\ref{aaa4})  we obtain \footnote{Actually, while rederiving  Eqs. (\ref{aaa46}), (\ref{aaa5}) in 2017 we have found additional multiplier $1/2\pi$, but since we were already quite sloppy with the numerical multiplier in going from summation to integration in Eq. (\ref{abr2}), we decided to leave the equations in this modified version as they were in the published version.}
\begin{eqnarray}
\label{aaa46}
&&\chi^{AA}\left({\bf R}_{ij}\right)=\frac{a^4}{256v_FR^3}\left[1+\cos(({\bf K}-{\bf K'}){\bf \cdot R}_{ij})\right]\\
\label{aaa5}
&&\chi^{AB}\left({\bf R}_{ij}\right)=-\frac{3a^4}{256v_FR^3}\left[1-\cos(({\bf K}-{\bf K'}){\bf \cdot R}_{ij}-2\theta_{\bf R})\right].\nonumber\\ \end{eqnarray}

The approach presented above  can be easily applied to the   bilayer graphene. We'll consider Bernal ($\tilde{A}-B$) stacking.  Because the low--energy modes are localized on $A$ and $\tilde{B}$ sites \cite{mccann}, we consider RKKY interaction of the magnetic adatoms siting on top of carbon atom in $A$ and/or $\tilde{B}$ sites. The low--energy modes are characterized by the spectrum \begin{eqnarray} E_{\pm}({\bf k})=\pm\frac{k^2}{2m} \end{eqnarray} and wave functions \begin{eqnarray} \psi_{\nu,{\bf K}}({\bf k})=\frac{1}{\sqrt{2}}\left(\begin{array}{l}e^{-i\theta_{\bf k}}\\ \nu e^{i\theta_{\bf k}}\end{array}\right)\nonumber\\ \psi_{\nu,{\bf K}'}({\bf k})=\frac{1}{\sqrt{2}}\left(\begin{array}{l}e^{i\theta_{\bf k}}\\ \nu e^{-i\theta_{\bf k}}\end{array}\right), \end{eqnarray} where this time the upper line of the spinor refers to the sublattice $A$ and the lower line refers to the sublattice $\tilde{B}$ \cite{mccann} (we ignore the trigonal warping). So for the case of bilayer we reproduce Eq. (\ref{rk}) (of course, the result for ${\cal G}^{AA}$ equally refers to ${\cal G}^{\tilde{B}\tilde{B}}$); Eq. (\ref{rk2}) is changed to \begin{eqnarray} \label{rk22} &&{\cal G}^{A\tilde{B}}(i,j;\tau>0)=\frac{1}{2}\frac{a^2}{(2\pi)^2}\int d^2{\bf k} e^{-E_+({\bf k})\tau}\nonumber\\ &&\times\left[ e^{i({\bf K}+{\bf k}){\bf \cdot R}_{ij}- 2i\theta_k}-e^{i({\bf K}'+{\bf k}){\bf \cdot R}_{ij}+ 2i\theta_k}\right]. \end{eqnarray} Calculation of ${\cal G}^{AA}$
 would demand  the integral \cite{watson}
\begin{eqnarray}
\int_0^{\infty}J_0(x)\exp(-px^2)xdx=\frac{1}{2p}\exp\left(-\frac{1}{4p}\right).
\end{eqnarray}
After simple calculus we obtain for bilayer graphene
\begin{eqnarray}
\label{aaa42}
\chi^{AA}\left({\bf R}_{ij}\right)=\frac{ma^4}{16\pi^2R^2}\left[1+\cos(({\bf K}-{\bf K'}){\bf \cdot R}_{ij})\right].
\end{eqnarray}

We return to monolayer graphene. The case of magnetic adatom siting on top of carbon atom,  certainly does not exhaust all the possibilities for the adatom  positions in graphene lattice \cite{uchoa,cheianov}. However, under rather general assumptions the specific  position of the adatom can be taken into account by changing in Eq. (\ref{abr2}) the product of the components of the spinor wave function $\psi_{\bf p}$ to an appropriate matrix element. Thus, using the results of Ref. \cite{uchoa}, for the case of substitutional impurities instead of Eqs. (\ref{rk}) and (\ref{rk2}) we obtain
\begin{eqnarray}
\label{rkb}
&&{\cal G}^{AA}(i,j;\tau>0)=-\frac{1}{2}\frac{a^4}{(2\pi)^2}\int d^2{\bf k}k^2 e^{i{\bf k \cdot R}_{ij}-E_+({\bf k})\tau}\nonumber\\
&&\left[e^{i{\bf K \cdot R}_{ij}}+e^{i{\bf K' \cdot R}_{ij}}\right]\nonumber\\
&&{\cal G}^{AB}(i,j;\tau>0)=\frac{1}{2}\frac{a^4}{(2\pi)^2}\int d^2{\bf k}k^2 e^{-E_+({\bf k})\tau}\nonumber\\
&&\left[ e^{i({\bf K}+{\bf k}){\bf \cdot R}_{ij}- 3i\theta_k}-e^{i({\bf K}'-{\bf k}){\bf \cdot R}_{ij}+ 3i\theta_k}\right].
\end{eqnarray}
After simple calculus we obtain
\begin{eqnarray}
\label{aaa4b}
\chi^{S_AS_A}\left({\bf R}_{ij}\right)=\frac{X^{S_AS_A}}{v_FR^7}\left[1+\cos(({\bf K}-{\bf K'}){\bf \cdot R}_{ij})\right] \nonumber
\end{eqnarray}
and
\begin{eqnarray}
\label{aaa5bg}
\chi^{S_AS_B}\left({\bf R}_{ij}\right)=-\frac{X^{S_AS_B}}{v_FR^7}\left[1-\cos(({\bf K}-{\bf K'}){\bf \cdot R}_{ij}-6\theta_{\bf R})\right],\nonumber \end{eqnarray}
where $X^{S_AS_A}$ and $X^{S_AS_B}$ can be easily calculated analytically.

\section{Discussion}

In this Section  we would like to compare our results with the previously obtained ones and additionally justify our line of reasoning.

The correction to the thermodynamic potential can be also written down using frequency representation  \cite{abrikosov}, which would  give \begin{eqnarray} \label{effective} \chi_{ij}=\frac{T}{4}\sum_{n,m}u_n^*(i)u_m(i)u_n(j)u_m^*(j)\nonumber\\ \sum_{\omega}\frac{1}{i\omega-\xi_n}\frac{1}{i\omega-\xi_m}, \end{eqnarray} where $\omega=\pi T(2l+1)$ ($l$ is an integer) is Matsubara frequency. (Eq. (\ref{abr2}) was taken into account.) Performing in Eq. (\ref{effective}) summation with respect to Matsubara frequencies we obtain \begin{eqnarray} \label{effective6} \chi_{ij}=\frac{1}{4}\sum_{n,m}u_n^*(i)u_m(i)u_n(j)u_m^*(j) \frac{n_F(\xi_m)-n_F(\xi_n)}{\xi_n-\xi_m}. \end{eqnarray} This is in fact  the result  obtained  originally \cite{ruderman,kasuya,yosida}, by using standard quantum mechanics (off-the-energy shell) perturbation theory.

We used the term bipartite lattice, but actually neither the symmetry of spectrum , presented after Eq. (\ref{gan2}), nor the symmetry of wave functions presented in Eq. (\ref{pm}) do not require any space periodicity (or any order at all) in the position of the sites. They even do not require that the Hamiltonian will be Hermitian, so they remain, say, in non--Hermitian quantum mechanics.

The results of Ref. \cite{bunder}  correspond to Eq. (\ref{effective6}) with a small but substantial difference: the terms with $\xi_n=\xi_m$ are discarded \cite{bunder2}, which breaks the symmetry of the RKKY interaction we discussed. We want now to consider a simple toy model to additionally explain that these diagonal terms are relevant and should be where they are. Our arguments will follow the consideration of the magnetism of electron gas in Ref. \cite{landau}.

Let the spectrum of $H$ consists of pairs of states having the same energy, and $H_{int}$ has non-zero matrix elements only between the states belonging to the same pair. Then the quantum mechanical problem of finding the spectrum of the Hamiltonian $H_T$ can be solved exactly, each doublet is split, $E_n^{(1,2)}= E_n\pm |V_{n,1;n,2}|$. The thermodynamic potential is \begin{eqnarray} \Omega=\sum _{n,\pm}\Omega^{0}\left(E_n\pm|V_{n,1;n,2}|\right), \end{eqnarray} where $\Omega^{0}(E)$ is the thermodynamic potential of the isolated level with the energy $E$. Expanding with respect to interaction we obtain \begin{eqnarray} \Delta\Omega=\sum _n\frac{\partial^2\Omega^{0}}{\partial E_n^2}|V_{n,1;n,2}|^2 =-\sum _n\frac{\partial n_F(E_n)}{\partial E_n}|V_{n,1;n,2}|^2,\nonumber\\ \end{eqnarray} which  corresponds to the diagonal terms in  Eq. (\ref{effective6}). The issue of diagonal terms can be also connected to the difference between the real- and imaginary-time approaches the authors of Ref. \cite{bunder} emphasize in their paper. Our opinion is that calculation of magnetic susceptibility using real--time method (Kubo formula) gives the adiabatic susceptibility. On the other hand, for the calculation of the RKKY interaction  we need the isothermal susceptibility, which is given by the imaginary--time method.

Analytical calculations of the RKKY interaction  can be done using Eq. (\ref{effective6}). In this case instead of Eqs. (\ref{aaa46}) and (\ref{aaa5})  we would obtain \begin{eqnarray} \label{aaa4bf} &&\chi^{AA}\left({\bf R}_{ij}\right)=\frac{a^4}{4\pi^2v_FR^3}\left[1+\cos(({\bf K}-{\bf K'}){\bf \cdot R}_{ij})\right]\nonumber\\ &&\int_0^{\infty}dx xJ_0(x)\int_0^{\infty}dx'x'J_0(x')\frac{1}{x+x'}\\ \label{aaa5bf} &&\chi^{AB}\left({\bf R}_{ij}\right)=-\frac{a^4}{4\pi^2v_FR^3}\left[1-\cos(({\bf K}-{\bf K'}){\bf \cdot R}_{ij}-2\theta_{\bf R})\right]\nonumber\\ &&\int_0^{\infty}dx x J_1(x)\int_0^{\infty}dx'x'J_1(x')\frac{1}{x+x'}. \end{eqnarray} Eqs. (\ref{aaa4bf}) and (\ref{aaa5bf}) are particularly convenient to  be compared with the results of Ref. \cite{sherafati}. Using the identity \cite{prudnikov} \begin{eqnarray} \int_0^{\infty}\frac{x^{\nu}}{x+z}J_{\nu}(cx)dx=\frac{\pi z^{\nu}}{2\cos\nu\pi}[{\bf H}_{-\nu}(cz)-Y_{-\nu}(cz)],\nonumber\\ \end{eqnarray} where ${\bf H}_{\nu}(z)$ is the Struve function and $Y_{\nu}(z)$ is the Neumann function, we can present integrals in Eqs. (\ref{aaa4bf}) and (\ref{aaa5bf}) as \begin{eqnarray} \label{ii} \frac{\pi}{2}\int_0^{\infty}dx x^2 J_0(x)\left[Y_{0}(x)-{\bf H}_{0}(x)+\frac{2}{\pi x}\right]\nonumber\\ \frac{\pi}{2}\int_0^{\infty}dx x^2 J_1(x)[Y_{-1}(x)-{\bf H}_{-1}(x)]. \end{eqnarray} These integrals   are  similar to those standing in Eqs. (18) and  (25) of Ref. \cite{sherafati}, but contrary to the latter, our integrals diverge. This is guaranteed by the asymptotics of  Struve functions \begin{eqnarray} {\bf H}_{\nu}(x)-Y_{\nu}(x)\to\frac{1}{\sqrt{\pi}\Gamma\left(\nu+\frac{1}{2}\right)}\left(\frac{x}{2}\right)^{\nu-1}+O\left((x/2)^{\nu-3}\right).\nonumber\\ \end{eqnarray}

A deficiency of the previous analytic calculations of the RKKY interaction in graphene is, to our mind, not due to them using the frequency representation of the Green's function (though we find the imaginary time representation more convenient for the calculations), but due to them
 first calculating static spin susceptibility in momentum space
\begin{eqnarray} \label{fou0} \chi({\bf q})=\sum_{\nu\nu',{\bf p}}{\cal M}_{\nu,\nu',{\bf p},{\bf q}}\frac{n_F\left[E_{\nu'}({\bf p}+{\bf q})\right]-n_F\left[E_{\nu}({\bf p})\right]}{E_{\nu}({\bf p})-E_{\nu'}({\bf p}+{\bf q})} \end{eqnarray} (we shouldn't worry here what the matrix element ${\cal M}$ is) and then calculating  $\chi\left({\bf R}_{ij}\right)$
 making a Fourier transformation
\begin{eqnarray} \label{fou} \chi\left({\bf R}_{ij}\right)=\frac{a^2}{(2\pi)^2}\int d^2{\bf q}\chi\left({\bf q}\right)e^{i{\bf q\cdot R}_{ij}}. \end{eqnarray} Both integrals  turn out to be ultra--violet divergent, and cut-offs  should be introduced. We, on the other hand, calculated directly $\chi$ in real space representation, thus avoiding the problem of divergence of the integrals completely.

There is another   problem with calculating the RKKY interaction (in normal metals) which has a long history \cite{yafet,litvinov};  it arises when we combine the integrals (\ref{fou0}) and (\ref{fou}) into a single double integrals. The problem is which integration: with respect to ${\bf q}$ or with respect to ${\bf p}$ we should do first. We also avoid this problem completely.

The contact exchange interaction we used can be easily justified in the case of $s$--wave  orbital of the magnetic adatom \cite{uchoa}. The case of $d$--wave orbitals is more complicated. To find the physically meaningful form of Kondo perturbation, it is appropriate to go back to the possible origin of the Kondo model, i.e., the Anderson model. Following seminal paper by Schrieffer \cite{schrieffer}, let us specify the magnetic impurity as being the $S$--state ion, say Mn$^{++}$, whose $d$--shell has the configuration $S_{5/2}$. Since the $S$-state ion cannot change the orbital angular momentum of a conduction electron, one should use states which transform according to the irreducible representations of the point group of the crystal about the impurity center \cite{schrieffer}.

Such approach for the case of $d$--wave orbitals was realized  by Zhu et al. \cite{zhu} (see also Ref. \cite{uchoa2}). Considering the magnetic impurity above the center of the honeycomb (plaquette impurity), they started from the classification of the degenerate $3d$--orbitals of the magnetic atom  with respect to irreducible representations of the symmetry group $C_{6v}$ and inferred that $d_{z^2}$ belongs to  $A_1$,  $(d_{xz},d_{yz})$ belong  to $E_1$ and $(d_{x^2-y^2},d_{xy})$ belong  to $E_2$ representations. Specifying their approach, we'll  take into account hybridization of the $d$--orbitals of the magnetic impurity  with the $p^z$ states of the carbon atoms around the plaquette. The selection rules for matrix elements demand that from the states $|i>$, where $i\in {\cal P}$, and ${\cal P}$ is the set of sites surrounding the plaquette, we'll chose combinations realizing the same representations as above. Thus the hybridization Hamiltonian for the $3d$ magnetic impurity in terms of the irreducible reps of the system will take the form \begin{eqnarray} \label{hybrid} H_{hyb}=\sum_{\lambda,\alpha,i\in {\cal P}}\left(v^{\lambda}_{i}c^{\dagger}_{i\alpha}f_{\lambda\alpha} +h.c.\right), \end{eqnarray} where operators $f^{\dagger}$ ($f$) create (annihilate) electrons at the $d$--orbitals of the magnetic impurity, and index $\lambda$ enumerates the orbitals $d_{z^2},d_{xz},d_{yz},d_{x^2-y^2},d_{xy}$. From Eq. (\ref{hybrid}), following Ref. \cite{schrieffer} under appropriate assumptions we can get the  $p-d$ exchange model \cite{uchoa} \begin{eqnarray} \label{pd} H_{pd}=-\sum_{\lambda,\alpha,\beta,i,j\in {\cal P}}Jv^{\lambda}_{i}{v^{\lambda}_{j}}^*{\bf S\cdot\sigma}_{\alpha\beta}c^{\dagger}_{i\alpha}c_{j\beta}. \end{eqnarray}

In Ref. \cite{saremi}, the $p-d$ exchange Hamiltonian (for the so-called coherent case) was previously taken in a very specific form \begin{eqnarray} \label{pd2} H_{pd}=-J\sum_{\alpha,\beta,i,j\in {\cal P}}{\bf S\cdot\sigma}_{\alpha\beta}c^{\dagger}_{i\alpha}c_{j\beta}, \end{eqnarray}
 which in fact takes into account only
the hybridization between $d_{z^2}$ and the combination of the $p$--states on the plaquette, realizing  irreducible representation $A_1$, that is
 $\sum_{i\in {\cal P}}|i>/\sqrt{6}$. Such specific form led to the conclusion that $1/|{\bf R}-{\bf R}'|^3$ term in the RKKY interaction between the
plaquette impurities vanishes. When the general form of the hybridization Hamiltonian  (\ref{hybrid}) is taken into account, this conclusion seems to us unjustified.

I am  grateful to B. Uchoa, J. Bunder, I. Titvinidze,  M. Potthoff, and L. Sandratskii for very useful discussions.

The work was done during the author's visit to Cavendish Laboratory, Cambridge University and finalized during the author's visit to the I. Institute of Theoretical Physics, Hamburg University. Important additions were made during the author's visit to Max Planck Institute of Microstructure Physics, Halle.

\begin{appendix}
\section{[Added in 2017] Finite temperature}

Above we considered only the case $T=0$, which corresponded to infinite upper integration limit in Eq. (\ref{abr}).
However, consideration of finite temperature just modifies our previous results  in a simple way. Thus,
taking into account that
\begin{eqnarray}
&&{\cal G}^{AA}(i,j;\tau>0)=-\frac{a^2}{4\pi}\frac{v\tau}{\left(v^2\tau^2+R^2\right)^{3/2}}\nonumber\\
&&\left[e^{i{\bf K \cdot R}_{ij}}+e^{i{\bf K' \cdot R}_{ij}}\right]\\
&&{\cal G}^{AB}(i,j;\tau>0)=\frac{a^2}{4\pi}\frac{R}{\left(v^2\tau^2+R^2\right)^{3/2}}\nonumber\\
&&\times\left[ e^{i({\bf K}+{\bf k}){\bf \cdot R}_{ij}- i\theta_k}-e^{i({\bf K}'+{\bf k}){\bf \cdot R}_{ij}+ i\theta_k}\right].
\end{eqnarray}
we obtain
\begin{eqnarray}
\label{new1}
&&\chi^{AA}_T\left({\bf R}_{ij}\right)=\chi^{AA}\left({\bf R}_{ij}\right)\frac{16}{\pi}\int_0^{v/RT}\frac{x^2dx}{(x^2+1)^3}\\
\label{new2}
&&\chi^{AB}_T\left({\bf R}_{ij}\right)=\chi^{AB}\left({\bf R}_{ij}\right)\frac{16}{3\pi}\int_0^{v/RT}\frac{dx}{(x^2+1)^3},\nonumber\\
\end{eqnarray}
where  $\chi^{AA}\left({\bf R}_{ij}\right)$ and $\chi^{AB}\left({\bf R}_{ij}\right)$ are given by Eqs. (\ref{aaa46}) and
(\ref{aaa5}) respectively.

Integrals in Eqs. (\ref{new1}), (\ref{new2}) can be easily calculated, but we'll restrict ourselves only by analyzing the limiting cases.
For $T\ll v/R$ we obtain the previous ($T=0$) results, in the opposite limiting case $T\gg v/R$ we get
\begin{eqnarray}
\label{new1b}
&&\chi^{AA}_T\left({\bf R}_{ij}\right)=\chi^{AA}\left({\bf R}_{ij}\right)\frac{16}{\pi}\left(\frac{v}{RT}\right)^3\\
\label{newb}
&&\chi^{AB}_T\left({\bf R}_{ij}\right)=\chi^{AB}\left({\bf R}_{ij}\right)\frac{16}{3\pi }\frac{v}{RT}.
\end{eqnarray}
We must mention that comparing our results with those obtained earlier for the case of doped graphene \cite{klier}, one should be aware of the fact that the exponential decrease of the RKKY interaction with the
distance at high temperatures  obtained in Ref. \onlinecite{klier}, was obtained for $k_FR\gg 1$ (in our case $k_F=0$).

This revision of our previous results was triggered by the author's short visit to the Physics Department of Bonn University and discussions with J. Kroha and  T. A. Costi, which are gratefully acknowledged, and was performed during the author's long visit to
Max-Planck-Institut fur Physik komplexer Systeme. The author  cordially thanks the Institute for the hospitality extended to him during
that and all the  previous visits.

\end{appendix}

\end{document}